\documentclass[aps,reprint,prr,twocolumn,floatfix,amsmath,amssymb]{revtex4}
\usepackage[pdftex]{graphicx} 
\usepackage{color}
\usepackage{dcolumn}
\usepackage{comment}
\begin{document}
\title{
Anomalous rheology of puller-type microswimmer suspensions
}
\author{Haruki Hayano}
\thanks{hayano@iis.u-tokyo.ac.jp}
\author{Akira Furukawa}
\thanks{furu@iis.u-tokyo.ac.jp}
\affiliation{Institute of Industrial Science, 
University of Tokyo, Meguro-ku, Tokyo 153-8505, Japan}
\date{\today}
\begin{abstract}
  We explore the mechanism underlying the anomalous rheology of puller-type microswimmer suspensions through direct hydrodynamic simulations. 
  Puller-type swimmers generate {\it contractile} flow fields along their swimming direction, leading to hydrodynamic interactions that cause the swimmers to align vertically. Our simulations reveal that this alignment effect, along with the resultant orientational order of swimming motion, becomes particularly pronounced near boundary {\it walls}, where local swimmer density is amplified, predominantly controlling the overall swimming dynamics and rheological properties of the suspension. These findings contrast with our previous simulations of pusher-type swimmers, which hydrodynamically interact through {\it extensile} flow fields, whereby they exhibit weak orientational order in the {\it bulk} region, which primarily determines their steady-state properties. 
  Furthermore, we demonstrate that the steady-state behavior near the walls is strongly influenced by the aspect ratio of the microswimmers and the degree of confinement between the walls.    
  Our results highlight the crucial role of microswimmer characteristics, such as shape and swimming mechanisms, in determining the rheological properties of active suspensions.
\end{abstract}

\maketitle

\section{Introduction}
Unique rheological properties, absent in passive suspensions, have been observed in active suspensions \cite{Sokolov,Gachelin,Lopez,Liu,PNAS2020,Chui,Xu,Rafai,Mussler,Review2,Marchetti,Review3}.
For instance, in suspensions of {\it E. coli}, categorized as pusher-type microswimmers, viscosity decreases sharply below the solvent viscosity at relatively low concentrations and shear rates \cite{Sokolov,Gachelin,Lopez,Liu,PNAS2020,Chui,Xu}, frequently resulting in a superfluid state with zero viscosity \cite{Lopez,PNAS2020,Chui}. 
In contrast, in suspensions of {\it Chlamydomonas reinhardtii}, classified as puller-type microswimmers, a steeper increase in viscosity compared to non-motile cells has been experimentally observed \cite{Rafai,Mussler}. 
Such anomalies in viscosity of active suspensions were first predicted by a seminal phenomenological model proposed by Hatwalne {\it et al.} \cite{Hatwalne}, suggesting that stress can vary due to the swimmers' self-propulsive forces when swimmers have orientational order under shear flow.

For rod-like particles, thermal or athermal rotational diffusion combined with shear flow produces weak orientational order \cite{Hinch_Leal1,Hinch_Leal2}, with which self-propulsive forces can provide nonzero active stress \cite{Marchetti,Review3,Choudhary,Lavigne}.
However, the substantial origin of rotational diffusion and orientational order in active suspensions remains elusive. 
One potential candidate is hydrodynamic interactions (HIs) among constituent swimmers, which may significantly contribute to the rotational diffusion of rod-like swimmers \cite{Ryan,Gyrya,Saintillan-ShelleyP}.
Recently, our hydrodynamic simulations \cite{Hayano} using model pusher-type swimmers designed to replicate {\it E. coli} suspensions reveal that, under shear flow, weak orientational order systematically evolves via hydrodynamic scattering during their travel between boundary walls.
The simulations also show that swimmers near the walls move in the forward-flow direction, which is an important characteristic of rod-like pusher-type swimmers \cite{Liu,Hill,Nash,Costanzo,Kaya,comment_flow}.

Meanwhile, as noted above, it has been experimentally observed that the viscosity of suspensions of motile {\it Chlamydomonas} is significantly larger than that of non-motile ones \cite{Rafai,Mussler}, which contrasts with the case of pusher suspensions. 
However, it is not trivial, even qualitatively, to determine what steady-state swimming properties are responsible for such anomalous rheology distinct from those observed for pusher-type swimmers.
Simulation studies by Jibuti {\it et al.} \cite{Jibuti}, in which a model swimmer with a spherical body and two beating spherical flagella replicating {\it Chlamydomonas} is used, show that due to the beating motions of flagella, the rotational motions of the swimmers are notably different from those of rod-like particles, leading to a significant increase in suspension viscosity. 
Although the simulation results and their physical implications are appealing, the potential influence of many-body HIs, which can greatly impact steady-state behaviors, should be further investigated. 
Additionally, in actual experimental systems, boundary walls exist, inevitably affecting the dynamics of swimmers and the flow field near walls. It is unclear how these effects influence overall rheological properties and how they differ from those of pusher-type swimmers.

In this paper, we numerically study the steady-state properties, particularly the rheological properties, of model puller swimmers through direct hydrodynamic simulations. The swimming mechanisms and the resultant flow fields generated around the swimmers differ significantly between pusher- and puller-type swimmers: For pusher swimmers, propulsion occurs through the generation of extensile forces, resulting in extensile flow fields along their swimming direction. In contrast, puller swimmers generate propulsion via contractile forces, leading to contractile flow fields along their moving direction. 
Since HIs between swimmers are mediated by these induced flow fields, the distinct flow characteristics of pushers and pullers are expected to play a crucial role in determining the rheological properties of their suspensions.
However, the precise mechanisms by which these differences emerge remain unclear. Furthermore, the steady-state swimming behaviors that accompany the experimentally observed anomalous rheology in puller-type microswimmer suspensions are not yet well understood.
To address these problems, this study aims to elucidate the physical mechanisms underlying the anomalous rheology of puller-type microswimmer suspensions.

The organization of this paper is as follows.
In Sec. II, we describe the details of our model swimmers and simulation setup. We also discuss the properties of the flow field generated by a single swimmer.
In Sec. III, supplemented by Appendix, we present simulation results for the steady swimming states and their corresponding rheological properties. Consistent with experimental observations, our model puller-type swimmers exhibit anomalous viscosity larger than that of passive-particle suspensions. We attribute this increase in viscosity to the enhanced orientational order of swimmers, induced by the contractile flow fields generated around individual swimmers. This effect is particularly pronounced near boundary walls, where swimmers tend to accumulate due to longer residence times. 
Furthermore, we demonstrate that these rheological behaviors and orientational orders strongly depend on the shape of the swimmers (variations in aspect ratio) and the system size (the distance between the confining walls).
Finally, in Sec. IV, we provide concluding remarks, summarizing the key findings and their physical implications.

\section{Model swimmer system}

\begin{figure*}[hbt] 
  \includegraphics[width=0.93\textwidth]{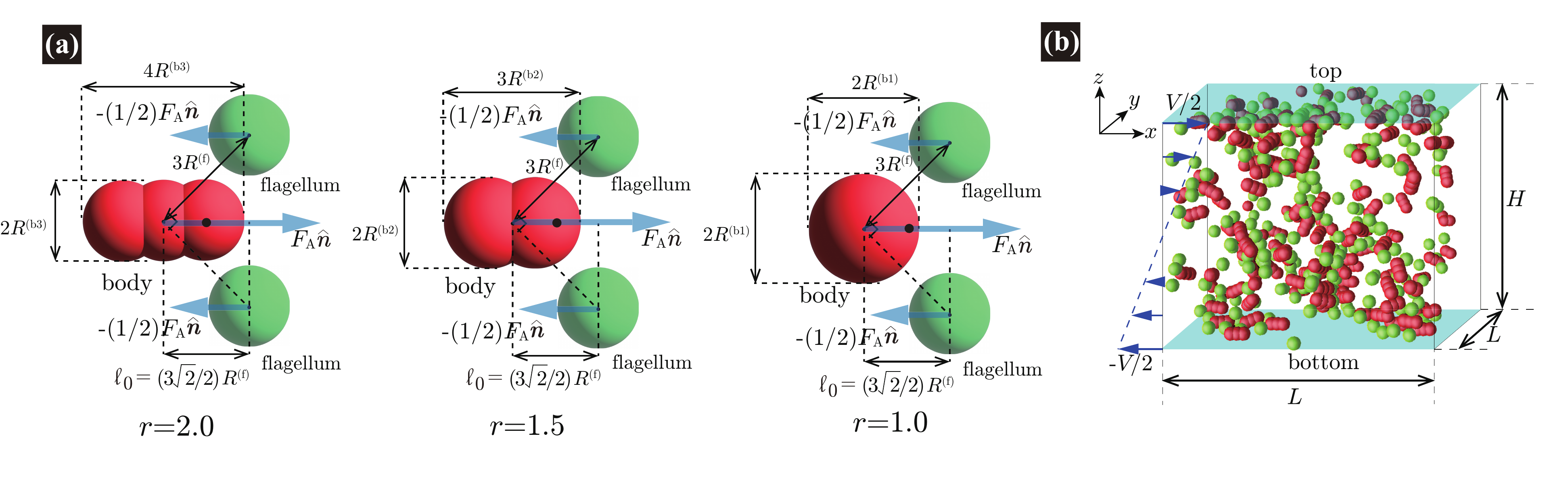}
  \caption{(Color online)
  (a)  Our model swimmers consist of a body and flagellum parts. 
  In this study, we consider three different aspect ratios of the body part, $r=1.0, 1.5$, and $2.0$, while keeping the body-part volume ${\mathcal V}^{\rm (b)}$ constant.
  For these three models, the body is constructed using different sphere configurations: 
  a three-sphere body with sphere radius $R^{\rm (b3)}$, a two-sphere body with sphere radius $R^{\rm (b2)}$, and a single-sphere body with sphere radius $R^{\rm (b1)}$.
  For the three-sphere body, the sphere centers are located at ${{\mbox{\boldmath$R$}}}^{\rm (G)}$, and ${{\mbox{\boldmath$R$}}}^{\rm (G)}\pm R^{\rm (b3)}{\hat {\mbox{\boldmath$n$}}}$, where ${{\mbox{\boldmath$R$}}}^{\rm (G)}$ is the swimmer's center-of-mass position, and ${\hat {\mbox{\boldmath$n$}}}$ is the unit vector representing its orientation. 
  For the two-sphere body, the sphere centers are located at ${{\mbox{\boldmath$R$}}}^{\rm (G)}\pm (R^{\rm (b2)}/2){\hat {\mbox{\boldmath$n$}}}$. 
  For the single-sphere body, the sphere center is at ${{\mbox{\boldmath$R$}}}^{\rm (G)}$. 
  On the other hand, the flagellum part is composed of two spheres of radius $R^{\rm (f)} (= R^{({\rm b3})})$. 
  The centers of these spheres are located at 
  ${{\mbox{\boldmath$R$}}}_{i}^{\rm (f)}={{\mbox{\boldmath$R$}}}^{\rm (G)}+3R^{\rm (f)} {\hat {\mbox{\boldmath$t$}}}^{\rm (f)}_{i}$ ($i=1,2$). 
  In our model, the unit vectors ${\hat {\mbox{\boldmath$t$}}}^{\rm (f)}_{i}$ satisfy $({\hat {\mbox{\boldmath$t$}}}^{\rm (f)}_{1} + {\hat {\mbox{\boldmath$t$}}}^{\rm (f)}_{2})/\sqrt{2} = {\hat {\mbox{\boldmath$n$}}}$ and the orthogonality condition ${\hat {\mbox{\boldmath$t$}}}^{\rm (f)}_{1} \cdot {\hat {\mbox{\boldmath$t$}}}^{\rm (f)}_{2} = 0$. 
  An active force $F_{\rm A}{\hat {\mbox{\boldmath$n$}}}$ is exerted on the body, while $-(F_{\rm A}/2){\hat {\mbox{\boldmath$n$}}}$ is applied directly to the solvent through each of the two flagellum spheres.
  At a sufficient distance from the swimmer, this swimmer behaves as a force dipole with a dipole moment magnitude of $F_{\rm A} \ell_0$, located at the point represented by the black circle.
  Here, $\ell_0$ is the characteristic swimmer length, given by $\ell_0 = (3 \sqrt{2}/2) R^{\rm (f)}$ for the present model.  
  (b) Our model suspension imposes periodic boundary conditions in the $x$- and $y$-directions, with the linear dimension $L$. Shear flow is imposed by moving the top and bottom walls in the $x$-direction at constant velocities $V/2$ and $-V/2$, respectively. These two walls are separated in the $z$-direction by a distance $H$, giving rise to a mean shear rate of $V/H$.
  \label{fig:model_pullers2}
  } 
\end{figure*}

\begin{figure}[hbt] 
  \includegraphics[width=0.48\textwidth]{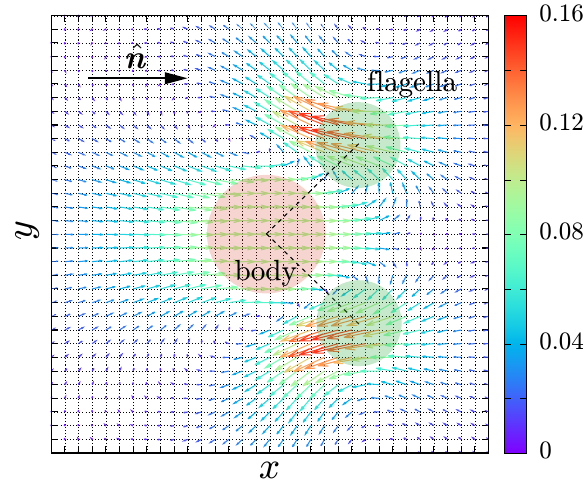}
  \caption{(Color online) 
  The steady-state velocity field $\hat{\mbox{\boldmath$v$}}(x,y,z=0)$ generated around the present model puller-type swimmer with a spherical body ($r=1.0$) and two flagellum parts. The swimmer's orientation axis $\hat{\mbox{\boldmath$n$}}$ lies in the plane. The color bar indicates the magnitude of the velocity field. 
  The swimmer generates a contractile flow field along its swimming direction. The flagellum parts pull the surrounding fluid inward, while the body pushes the fluid forward, leading to a flow field characteristic of a force dipole at a sufficient distance from the swimmer.}
  \label{fig:velocity_3d_puller}
\end{figure}

In this study, we numerically investigate the current problems through direct hydrodynamic simulations of model puller-type microswimmers.
Our model swimmer consists of a rigid body and two flagellum-like parts. The body part is treated as a rigid particle, while the flagellum parts are represented by two massless spherical particles that simply follow the body's motion. This treatment ensures that the relative positions of the body and flagellum components remain unchanged.
For the $\alpha$-th swimmer ($\alpha=1,\cdots, N$), where $N$ is the total number of swimmers, we assume that a force $F_{\rm A}{\hat {\mbox{\boldmath$n$}}}_{\alpha}$ acting on the body is exerted by the flagellum parts.
Each flagellum component, in turn, applies a force of $-(1/2)F_{\rm A}{\hat {\mbox{\boldmath$n$}}}_{\alpha}$ directly to the solvent fluid. 
Here, ${\hat {\mbox{\boldmath$n$}}}_{\alpha}$ represents the direction of the $\alpha$-th swimmer. The present particle-based model, schematically shown in Fig. \ref{fig:model_pullers2}(a), is essentially the same as those proposed in Refs. \cite{Graham1,Graham2} and used in subsequent studies \cite{Haines,Ryan,Gyrya,Decoene,Furukawa_Marenduzzo_Cates,Hayano,Kanazawa}.
Notably, while most of these earlier studies focused on pusher-type swimmers, the current simulations are conducted for puller-type microswimmers.
  
In Fig. \ref{fig:model_pullers2}(a), we illustrate the details of constructions of our model swimmers. 
To address the effects of swimmer shape on rheological properties, we considered three different aspect ratios $r=1.0, 1.5$, and $2.0$, while keeping the body-part volume ${\mathcal V}^{\rm (b)}$. 
For these three models, the body part is constructed using different sphere superpositions: 
a three-sphere configuration with radius $R^{\rm (b3)}$, a two-sphere configuration with radius $R^{\rm (b2)}$, and a single sphere configuration with radius $R^{\rm (b1)}$. 
In all three models, the flagellum components are represented by two identical spheres of radius $R^{\rm (f)} (= R^{\rm (b3)})$.

As mentioned above, an active force $F_{\rm A}{\hat {\mbox{\boldmath$n$}}}$ is exerted on the body, while a force of $-(F_{\rm A}/2){\hat {\mbox{\boldmath$n$}}}$ is directly exerted on the solvent through each of the two flagellum parts, where ${\hat {\mbox{\boldmath$n$}}}$ represents the swimmer's orientation. 
At a sufficient distance from the swimmer, the swimmer behaves as a force dipole with a magnitude of $F_{\rm A} \ell_0$, centered at the black circle indicated in Fig. \ref{fig:model_pullers2}(a). 
Here, $\ell_0$ represents the characteristic swimmer length, given by $\ell_0 = (3 \sqrt{2}/2) R^{\rm (f)}$ for the present model. 
    
The steady-state velocity field generated around a single swimmer is shown in Fig. \ref{fig:velocity_3d_puller} for the case of a spherical body ($r=1.0$). The swimmer induces a contractile flow field along its swimming direction, resembling the experimentally observed flow field of {\it Chlamydomonas} \cite{Drescher}: 
The flagellum parts pull the surrounding fluid inward, while the body pushes the fluid forward. 
As demonstrated in the following section, this contractile flow field leads to attractive HIs between the present puller-type swimmers along the vertical direction of their swimming orientations, enhancing the polar orientational order of the system.
  
As illustrated in Fig. \ref{fig:model_pullers2}(b), periodic boundary conditions are imposed in the $x$- and $y$-directions with the linear dimension $L$, and the planar top and bottom walls are placed at $z=H/2$ and $-H/2$, respectively. The shear flow is imposed by moving the top and bottom walls in the $x$-direction at constant velocities of $V/2$ and $-V/2$, respectively, whereby the mean shear rate is $\dot\gamma=V/H$. 
In our simulations, HIs between swimmers are incorporated using the Smoothed Profile Method (SPM) \cite{SPM,SPM2,SPM3}, which is one of the mesoscopic simulation techniques \cite{LB1,LB2,FPD,FPD2,MPC,DPD}. 
Among these approaches, SPM and Fluid-Particle Dynamics (FPD) explicitly solve the Navier-Stokes equations for solvent dynamics, providing two crucial advantages: (i) direct treatment of solvent degrees of freedom and (ii) reduced computational cost by approximating rigid solid-liquid boundary conditions using smoothed interfaces. In SPM, particle rigidity is enforced via a constraint force \cite{SPM,SPM2,SPM3}, allowing for larger time steps, which is an advantage over FPD, where smaller time steps are required to maintain numerical stability. However, while FPD accurately preserves energy conservation \cite{FPD2}, SPM does not inherently ensure this condition. Further details of the model and simulations are provided in the Supplementary Information.

\section{Results and discussion}

\subsection{Steady-state properties: Orientational order near the walls and contribution to the viscosity}

\begin{figure}[bh] 
  \includegraphics[width=0.48\textwidth]{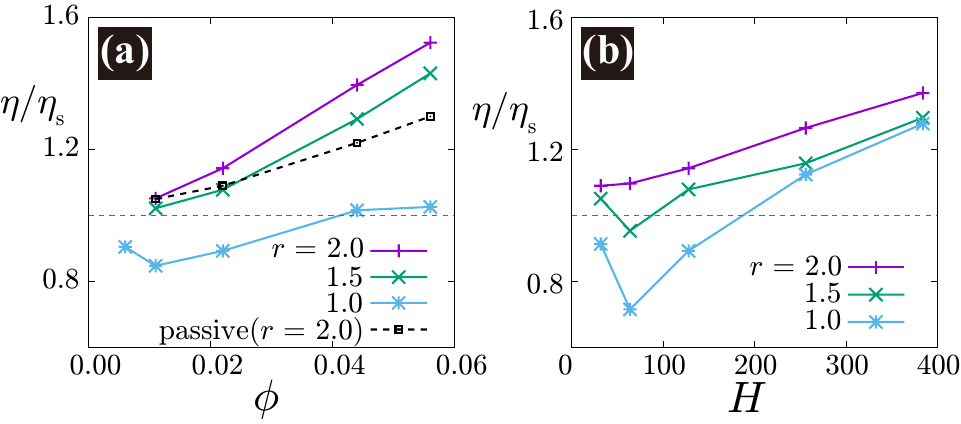}
  \caption{(Color online)   
  (a) The normalized viscosity $\eta/\eta_{\rm s}$ as a function of $\phi$ for various aspect ratios $r$ at $H=128$ and $\dot\gamma=10^{-3}$. 
  For rod-like swimmers ($r=2.0$ and $1.5$), the viscosity is higher than the solvent viscosity for all $\phi$.  For $r=2.0$, passive suspensions ($F_{\rm A}=0$) exhibit lower viscosities than active suspensions, indicating activity-induced viscosity enhancement. 
  In contrast, for spherical swimmers ($r=1.0$), the viscosity is lower than the solvent viscosity for $\phi\lesssim 0.04$. 
  (b) The normalized viscosity $\eta/\eta_{\rm s}$ as a function of $H$ for $\phi = 0.022$ and $\dot\gamma=10^{-3}$. 
 The viscosity generally increases with $H$, except at small $H$. In both (a) and (b), the dashed lines represent  $\eta/\eta_{\rm s}=1$. 
  }
  \label{fig:viscosity}
\end{figure}

First, Fig. \ref{fig:viscosity} presents the normalized viscosity $\eta/\eta_{\rm s}$ under various conditions.
In this study, the viscosity is defined as 
\begin{eqnarray}
\eta=\dfrac{1}{{\dot \gamma}L^2}\int {\rm d}x{\rm d}y\langle \Sigma_{xz}(x,y,\pm H/2)\rangle,
\label{eq:def_viscosity}
\end{eqnarray} 
where $\Sigma_{xz}(x,y,\pm H/2)$ is the $xz$ component of the stress tensor at the walls and $\langle \cdots \rangle$ hereafter denotes the time average in the steady-state. 
We find that $\eta$ increases with increasing $\phi$ for all aspect ratios $r$. 
Here, the volume fraction of the swimmers is defined as $\phi=N{\mathcal V}^{\rm (b)}/L^2H$.

Figure \ref{fig:viscosity}(a) shows that, for rod-like swimmers ($r=2, 1.5$) at $H=128$, the viscosity is higher than the solvent viscosity for all $\phi$.
For $r=2.0$, we also show the results for passive particles, where the active force is set to zero, $F_{\rm A}=0$, while keeping all other conditions unchanged, showing that passive suspensions exhibit lower viscosities than those for the active suspensions.
This result is consistent with experimental observations of {\it Chlamydomonas} suspensions \cite{Rafai, Mussler}, where the viscosity of motile living cells was found to be higher than that of non-motile dead cells, demonstrating activity-induced viscosity enhancement.
However, for spherical swimmers ($r=1.0$), the viscosity is lower than the solvent viscosity for $\phi\lesssim 0.04$ at this specific confinement height ($H=128$).
To examine whether this difference in the viscosity depends on $H$, in Fig. \ref{fig:viscosity}(b), we plot $\eta/\eta_{\rm s}$ for all $r$ as a function of $H$.
This plot reveals a significant dependence of the viscosity on both $H$ and $r$; 
the viscosity generally increases as increasing $H$, except at smaller $H(\lesssim 64)$.  
For the spherical swimmers, the viscosity is significantly higher than the solvent viscosity at larger $H$. Furthermore, even for rod-like swimmers ($r=1.5$), the viscosity is lower than the solvent viscosity at $H=64$. 
As discussed below (especially, in Sec. IIIC), the observed reduction in the viscosity is attributed to strong confinement effects: the swimming states differ remarkably above and below a certain threshold value of $H$, which itself depends on the aspect ratio $r$ of the swimmers.  

The viscosity can be divided into three parts: $\eta=\eta_{\rm s}+\Delta\eta_{\rm p}+\Delta\eta_{\rm a}$, where $\eta_{\rm s}$ is the solvent viscosity, while $\Delta\eta_{\rm p}$ and $\Delta\eta_{\rm a}$ represent the contributions from passive stress and active stress, respectively.
In the present framework, $\Delta\eta_{\rm a}$ is given by
\begin{eqnarray}
\Delta\eta_{\rm a} = \dfrac{1}{{\dot\gamma}L^2 H} F_{\rm A} \ell_0\sum_{\alpha=1}^N  \langle \hat{n}_{\alpha,x}(t) \hat{n}_{\alpha,z}(t)\rangle, 
\label{eq:active_viscosity}
\end{eqnarray}
where ${\hat{\mbox{\boldmath$n$}}}_\alpha(t)$ is the unit vector representing the orientation of the $\alpha$-th swimmer at time $t$, and ${\hat{n}}_{\alpha,x}$ and ${\hat{n}}_{\alpha,z}$ are its $x$- and $z$-components, respectively. The derivation of Eq. (\ref{eq:active_viscosity}) is provided in Supplementary Information.  
This expression is general for swimmers characterized by a prescribed force dipole. Note that, for pusher-type swimmers, the expression remains the same except for a minus sign. 
Essentially identical expressions, aside from minor differences, have been derived in previous studies (see Refs. \cite{Haines, Saintillan1, Review3}, for example).
From Eq. (\ref{eq:active_viscosity}), when swimmers tend to align along the extension direction of the flow field (i.e., when $\langle \hat{n}_{\alpha,x} \hat{n}_{\alpha,z}\rangle>0$), the active contribution to the viscosity is positive ($\Delta\eta_{\rm a}>0$).

\begin{figure}[hbt] 
 \includegraphics[width=0.48\textwidth]{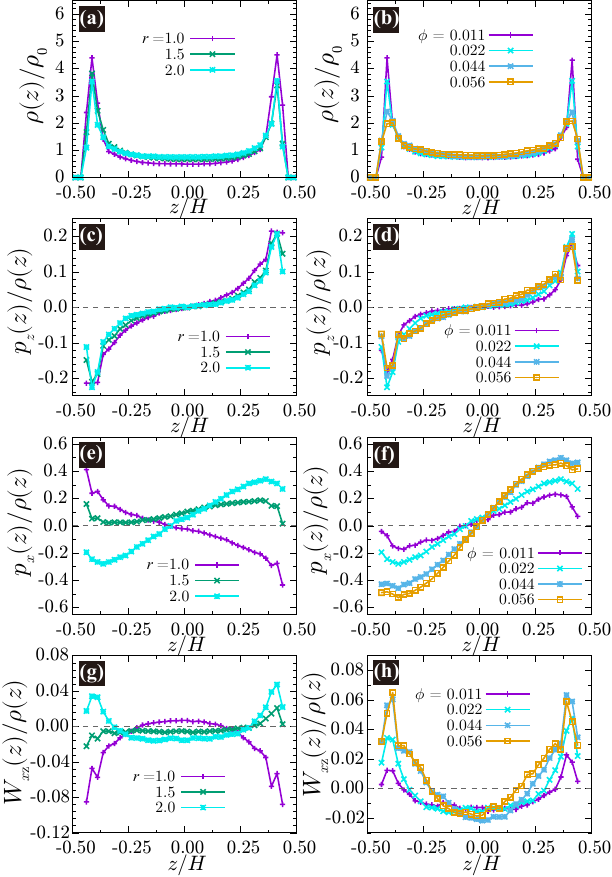}
  \caption{(Color online) 
  Subfigures (a), (c), (e), and (g) represent $\rho(z)/\rho_0$, $p_{z}(z)/\rho(z)$, $p_{x}(z)/\rho(z)$, and $W_{xz}(z)/\rho(z)$, respectively, for various aspect ratios $r$ at $\phi=0.022$, $\dot\gamma=10^{-3}$, and $H=128$. Here, $\rho_0 = N/(L^2 H)$ is the average number density.
  Subfigures (b), (d), (f), and (h) present the corresponding results for various $\phi$ at $r=2.0$, $\dot\gamma=10^{-3}$, and $H=128$.}
  \label{fig:8moments}
\end{figure}

To further investigate the swimming states involved in the anomalous viscosity, the following steady-state quantities were examined:  
${\rho}(z)=\sum_{\alpha=1}^{N} \langle \delta[{\mbox{\boldmath$r$}}-{\mbox{\boldmath$R$}_\alpha}(t)] \rangle$, ${\mbox{\boldmath$p$}}(z)=\sum_{\alpha=1}^{N} \langle {\hat{\mbox{\boldmath$n$}}}_\alpha (t) \delta[{\mbox{\boldmath$r$}}-{\mbox{\boldmath$R$}_\alpha}(t)]\rangle$, and ${\stackrel{\leftrightarrow}{\mbox{\boldmath$W$}}}(z)=\sum_{\alpha=1}^{N}\langle [{\hat{\mbox{\boldmath$n$}}}_\alpha(t){\hat{\mbox{\boldmath$n$}}}_\alpha(t)-{\stackrel{\leftrightarrow}{\mbox{\boldmath$\delta$}}}/3] \delta[{\mbox{\boldmath$r$}}-{\mbox{\boldmath$R$}_\alpha}(t)]\rangle $.
Here, ${\mbox{\boldmath$R$}_\alpha}(t) = \mbox{\boldmath$R$}_\alpha^{\rm (G)}(t)+(\ell_0/2) {\hat{\mbox{\boldmath$n$}}}_\alpha (t)$ is the center of the force dipole ($\ne$ the center-of-mass position), $\rho(z)$ is the density, and ${\mbox{\boldmath$p$}}(z)/\rho(z)$ and ${\stackrel{\leftrightarrow}{\mbox{\boldmath$W$}}}(z)/\rho(z)$ represent the polarization vector and the nematic order parameter tensor, respectively \cite{Review2}.  
These quantities, which depend only on $z$ at steady state, are shown in Figs. \ref{fig:8moments}(a)-(h) for various conditions.

In Figs. \ref{fig:8moments}(a) and (b), the normalized swimmer density $\rho(z)/\rho_0$ exhibits pronounced peaks near the boundary walls and remains nearly constant elsewhere, indicating that the walls effectively attract swimmers.
Similar behavior has been reported and discussed in previous studies (for example, Refs. \cite{Berke,Li-Tang,Review1,LaugaB,Figueroa-Morales,Bianchi,Denissenko,Ezhilan,Ezhilan-Saintillan,Yan-Brady}).
For $r=1.0$ and $1.5$, the results shown in Figs. \ref{fig:8moments}(a),(c),(e) and (g) are averaged over 12 independent runs due to symmetry breaking along the $z$-axis in individual runs. A detailed discussion of this effect is provided in Appendix A.

Figures \ref{fig:8moments}(c)-(f) display the polarization vector ${\mbox{\boldmath$p$}}(z)/{\rho}(z)$. 
The $z$-component, $p_z(z)/\rho(z)$, exhibits similar behavior regardless of the aspect ratio $r$ and concentration $\phi$. This indicates that swimmers moving near the walls tend to tilt their heads toward the walls, which causes them to remain close to the walls for a relatively longer time. This behavior has also been observed in previous studies on rod-like pusher swimmers (see, for example, Refs. \cite{Vigeant,Spagnolie_Lauga,Sipos}).
In contrast, the $x$-component, $p_x(z)/\rho(z)$, shows significant $r$-dependence:
For rod-like swimmers ($r=2.0$), $p_x(z)/\rho(z)$ is positive for $z>0$, and negative for $z<0$, suggesting that the swimmers move along the flow direction on average, which becomes most noticeable around the walls.
On the other hand, spherical swimmers ($r=1.0$) tend to move in directions opposite to the mean flow on average, which is also most pronounced around the walls. 
At $r=1.5$, a crossover between these two behaviors is observed; in the bulk region, $p_x(z)/\rho(z)$ shows nearly flat $z$-dependence, while the swimmers near the walls still exhibit a tendency similar to that at $r=1.0$.

Regarding viscosity enhancement or reduction, Figs. \ref{fig:8moments}(g) and (h) are of particular interest, as they show the $xz$-component of the nematic order parameter $W_{xz}(z)/\rho(z)$. 
For clarity, Eq. (\ref{eq:active_viscosity}) is rewritten as
\begin{eqnarray}
\Delta \eta_{\rm a} = \dfrac{1}{{\dot\gamma} H}F_{\rm A} \ell_0\int_{-H/2}^{H/2} {\rm d}z W_{xz}(z),
\label{eq:active_viscosity2}
\end{eqnarray}
which shows that $W_{xz}(z)$ is directly related to the viscosity enhancement or reduction. Distinct behaviors are observed between rod-like and spherical swimmers.
For rod-like swimmers ($r=2.0$), $W_{xz}(z)/\rho(z)$ is positive near the walls and slightly negative in the bulk region, and this trend becomes more pronounced with increasing $\phi$.
In contrast, for spherical swimmers ($r=1.0$), $W_{xz}(z)/\rho(z)$ is negative near the walls and positive in the bulk region.
These results, along with those in Figs. \ref{fig:8moments} (a) and (b), suggest that the behaviors of $W_{xz}(z)$ near the walls plays a dominant role in determining the viscosity enhancement (reduction) for the rod-like (spherical) swimmers. 
For $r=1.5$, an intermediate behavior is observed between the cases of $r=2.0$ and $r=1.0$.
However, as discussed in Sec. IIIC, the behaviors of $p_x(z)$ and $W_{xz}(z)$ for rod-like swimmers are also observed in spherical swimmers ($r=1.0$) at larger $H$. 
Note that the observed change in the viscosity is directly related to variations in the shear rate at the walls. A brief review of this behavior is provided in Appendix B. 

\subsection{Hydrodynamically induced vertical alignment of puller-type swimmers}

As noted in the previous subsection, the active-stress contribution to the overall viscosity, $\Delta \eta_{\rm a}$, is governed by Eq. (\ref{eq:active_viscosity}) or Eq. (\ref{eq:active_viscosity2}), where the $xz$-component of the nematic order parameter tensor $W_{xz}$ directly determines $\Delta \eta_{\rm a}$. While a similar relationship also holds for pusher-type swimmers, as investigated in our previous study \cite{Hayano}, a notable distinction emerges in the case of puller-type swimmers, as discussed below.

Due to the {\it contractile} nature of the active force dipole in puller swimmers, the resultant viscosity change is generally positive (viscosity enhancement), whereas for pusher swimmers, characterized by an {\it extensile} force dipole, the viscosity change is negative (viscosity reduction).
In Ref. \cite{Hayano}, we showed that for pusher swimmers, viscosity reduction is primarily determined by a significant positive value of $W_{xz}$ in the {\it bulk} region, while contributions from swimmers near the boundaries are minor. In contrast, for the present model puller swimmers, as clearly illustrated in Fig. \ref{fig:8moments}, the dominant contribution to $\Delta \eta_{\rm a}$ stems from swimmers near the {\it walls}, while the contribution from the bulk region remains relatively small.
This difference can be attributed to the enhanced polar order of puller swimmers, which tend to align vertically along their swimming directions. This alignment is especially pronounced near the walls. In this subsection, we analyze this effect in more detail.

To investigate correlations among swimmers near the walls, we compute the two-body correlation function measured along the $xy$-plane for swimmers located close to the walls, defined as:
\begin{flalign}
&g(\tilde{r},\tilde{\theta}) = & \nonumber \\ 
&\frac{L^2}{2 \tilde{r}N_{\rm w}(N_{\rm w}-1)} \sideset{}{^{\rm w}}{\sum}_{\alpha=1}^{N} \sideset{}{^{\rm w}}{\sum}_{\beta \ne \alpha}^{N} \langle \delta[\tilde{\theta} - \tilde{\theta}_{\alpha \beta}(t)]\delta[\tilde{r} - |\tilde{\mbox{\boldmath$r$}}_{\alpha \beta}(t)|]\rangle,
\end{flalign}
where $\tilde{\mbox{\boldmath$r$}}_{\alpha \beta}(t)$ represents the relative position vector between the $\alpha$-th and $\beta$-th swimmers in the $xy$-plane, given by $\tilde{\mbox{\boldmath$r$}}_{\alpha \beta}(t)= (X_{\beta}(t)-X_{\alpha}(t), Y_{\beta}(t)-Y_{\alpha}(t))$. 
Here, ${\mbox{\boldmath$R$}_\alpha}(t)=(X_\alpha(t),Y_\alpha(t),Z_\alpha(t))$. 
The relative orientation angle between the $\alpha$-th swimmer's orientation and the unit vector connecting the $\alpha$-th and $\beta$-th swimmers  is given by $\tilde{\theta}_{\alpha \beta} = \cos^{-1}({\hat{\mbox{\boldmath$n$}}}_{\alpha}^\bot \cdot \hat{\mbox{\boldmath$r$}}_{\alpha \beta})$, where $\hat{\mbox{\boldmath$r$}}_{\alpha \beta} = \tilde{\mbox{\boldmath$r$}}_{\alpha \beta}/|\tilde{\mbox{\boldmath$r$}}_{\alpha \beta}|$ is the unit vector in the direction of $\tilde{\mbox{\boldmath$r$}}_{\alpha \beta}$ and ${\hat{\mbox{\boldmath$n$}}}_{\alpha}^\bot=(\hat n_{\alpha,x}/\sqrt{ \hat n^2_{\alpha,x}+\hat n^2_{\alpha,y}},\hat n_{\alpha,y}/\sqrt{ \hat n^2_{\alpha,x}+\hat n^2_{\alpha,y}})$.
Furthermore, the summation $\sum^{\rm w}$ indicates that only swimmers near the walls are considered, those satisfying $|Z_{\alpha}/H|> 0.35$, and $N_{\rm w}$ is the average number of the swimmers meeting this condition.

\begin{figure}[hbt] 
  \includegraphics[width=0.48\textwidth]{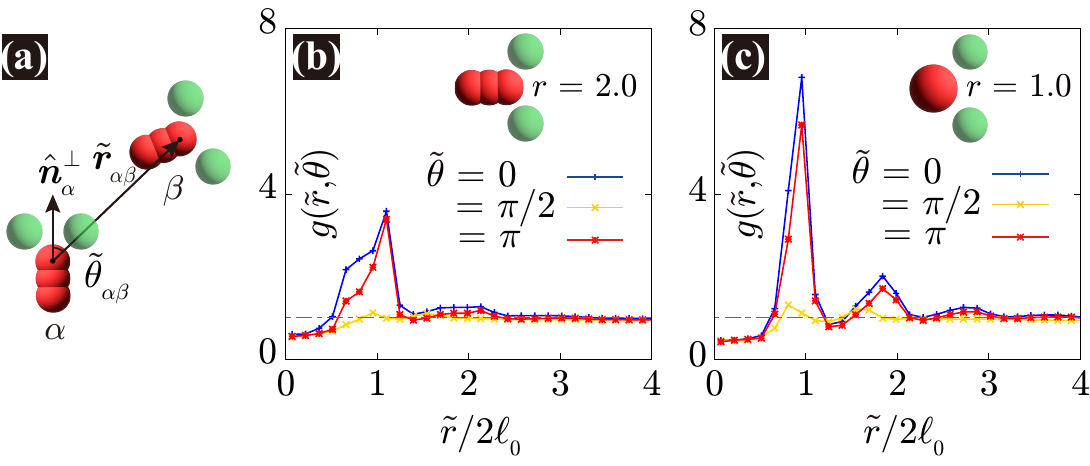}
  \caption{(Color online) (a) Schematic illustration of the definitions of $\tilde{\mbox{\boldmath$r$}}_{\alpha \beta}$ and ${\hat{\mbox{\boldmath$n$}}}_{\alpha}^\bot$. 
  The two-body correlation function $g(\tilde{r},\tilde{\theta})$ near the walls ($|z/H| > 0.35$) for $r=2.0$ (b) and $r=1.0$ (c) at $\phi=0.022$, $\dot\gamma=10^{-3}$ and $H=128$. In (b) and (c), the dashed lines indicate $g(\tilde{r},\tilde{\theta})=1$.  }
  \label{fig:gofr}
\end{figure}

In Fig. \ref{fig:gofr}, we plot $g(\tilde{r},\tilde{\theta})$ for three angular orientations: $\tilde{\theta}=0$, $\pi/2$, and $\pi$.
For both spherical and rod-like swimmers, at $\tilde{\theta}=0$ and $\pi$, $g(\tilde{r},\tilde{\theta})$ exhibits a distinct first peak at approximately $\tilde{r} =2\ell_0$. In contrast, for $g(\tilde{r},\pi/2)$, this peak is almost absent, indicating a strong vertical alignment tendency among the swimmers, which is a direct consequence of HIs.
As illustrated in Fig. \ref{fig:velocity_3d_puller}, the contractile active force prescribed in our model swimmer generates an inward flow along the swimmer's orientation. 
This flow induces attractive HIs among swimmers along their swimming directions, which manifests as the observed peak in $g(\tilde{r},\tilde{\theta})$ at $\tilde{\theta}=0$ and $\pi$.
Interestingly, the peaks for spherical swimmers are notably higher than those for rod-like swimmers, suggesting a stronger vertical alignment tendency for spherical swimmers. This difference may be attributed to steric effects: Let us consider two vertically aligned swimmers, where the body of one swimmer is positioned between the flagellum parts of another. 
This alignment configuration is likely to be more resistant to transverse fluctuations in spherical swimmers than in rod-like swimmers.
As demonstrated in Appendix A, the pronounced alignment tendency of spherical swimmers promotes the formation of larger, vertically aligned clusters, which can often extend to the system size ($L=128$).
It is important to note that this steric effect is significantly influenced by the specific structural setup of our model swimmers and may be somewhat artificial. 
However, the vertical alignment tendency itself arises from the characteristic contractile flow field of puller-type swimmers and is therefore expected to be a general feature in puller swimmer systems.
We further note that the present argument is not limited to swimmers near the walls but also applies to those in the bulk region. However, because the swimmer density near the walls is higher and their motion is more constrained along the surface, the alignment tendency is more pronounced in these regions than in the bulk. 

\subsection{Strong confinement effects on swimming states}

Thus far, we have focused on the steady-state swimming behaviors and their influence on rheological properties for a fixed system size ($H=L=128$).
Notably, we observe two distinct swimming states depending on the swimmers' aspect ratio. Rod-like swimmers ($r=2.0$) tend to follow the forward-flow direction ($\hat n_{\alpha,x} v_x>0$) on average, whereas spherical swimmers ($r=1.0$) exhibit an average swimming direction opposite to the mean flow ($\hat n_{\alpha,x} v_x<0$), where $v_x$ is the $x$-component of the solvent velocity.
In this subsection, we demonstrate that strong confinement effect is responsible for these contrasting behaviors.
As shown below, even spherical swimmers follow the forward-flow direction when $H$ is sufficiently large, while rod-like swimmers slightly align with the backward-flow direction when $H$ is smaller. The threshold value of $H$ that distinguishes these behaviors depends on the aspect ratio $r$, which determines the swimmers' sensitivity to applied torques.

\begin{figure}[htb]
  \includegraphics[width=0.48\textwidth]{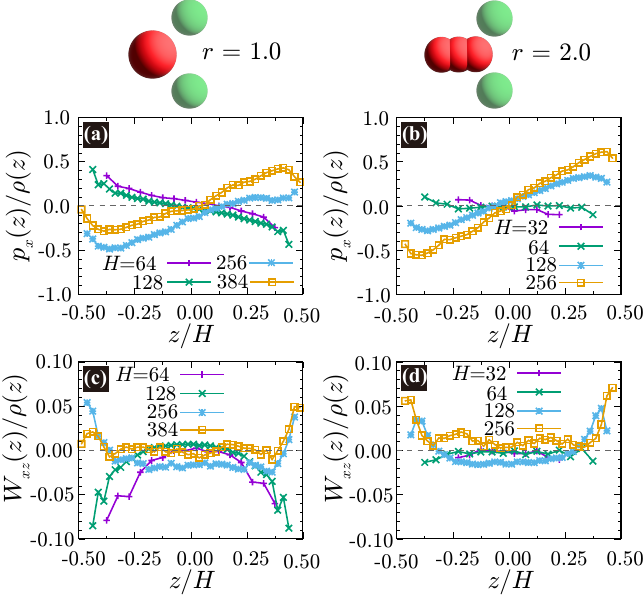}
  \caption{(Color online)
  The $H$-dependence of $p_x(z)$ and $W_{xz}(z)$ for the spherical($r=1.0$) and rod-like($r=2.0$) swimmers at $\phi=0.022$, $\dot\gamma=10^{-3}$ with various $H$.}
  \label{fig:H_dependance}
\end{figure}

We examine how the orientational states of both spherical ($r=1.0$) and rod-like ($r=2.0$) swimmers depend on $H$.
Figure \ref{fig:H_dependance} presents $p_x(z)/\rho(z)$ and $W_{xz}(z)/\rho(z)$ for spherical and rod-like swimmers at $\phi=0.022$ and $\dot\gamma=10^{-3}$ for various values of $H$.
The changes in $p_x(z)/\rho(z)$ are shown in Figs. \ref{fig:H_dependance}(a) and (b) for spherical and rod-like swimmers, respectively.
For spherical swimmers, although the average swimming direction is along the backward-flow direction at $H=64$ and $128$ (see Fig. \ref{fig:8moments}(e)), it reverses to the forward-flow direction at $H=256$ and $384$.
Therefore, a transition or crossover should occur between $H=128$ and $256$. 
Similarly, for rod-like swimmers, the average swimming direction is slightly aligned with the backward-flow direction at smaller values of $H$($=32$ and $64$), but it aligns with the forward-flow direction at $H=128$ and $256$.
We do not show $p_z(z)/\rho(z)$ here because it behaves qualitatively the same regardless of $H$ and $r$ (see  Fig. \ref{fig:8moments}(c) for $H=128$).
Reflecting the change in $p_x(z)$, $W_{xz}(z)/\rho(z)$ near the boundaries also changes sign, as shown in Figs. \ref{fig:H_dependance}(c) and (d).
Specifically, for the spherical swimmers, it is significantly negative at $H=64$ and $128$ but becomes positive at $H=256$ and $384$, whereas for rod-like swimmers it is slightly negative at $H=32$ and $64$, but positive at $H=128$ and $256$.
Recall that $W_{xz}(z)$ is directly related to the active-stress contribution to the viscosity, $\Delta \eta_{\rm a}$, through Eq. (\ref{eq:active_viscosity2}).
Consequently, for sufficiently large $H$, $\Delta \eta_{\rm a}$ should become positive for both spherical and rod-like swimmers.
This result is consistent with the experimental observations in {\it Chlamydomonas} suspensions, where motile (active) cells exhibit a significantly higher viscosity than immotile (passive) cells. 
Thus, we expect that the orientational orders found here (particularly those pronounced near the walls) account for such experimental findings. 

\begin{figure}[bth]
  \includegraphics[width=0.48\textwidth]{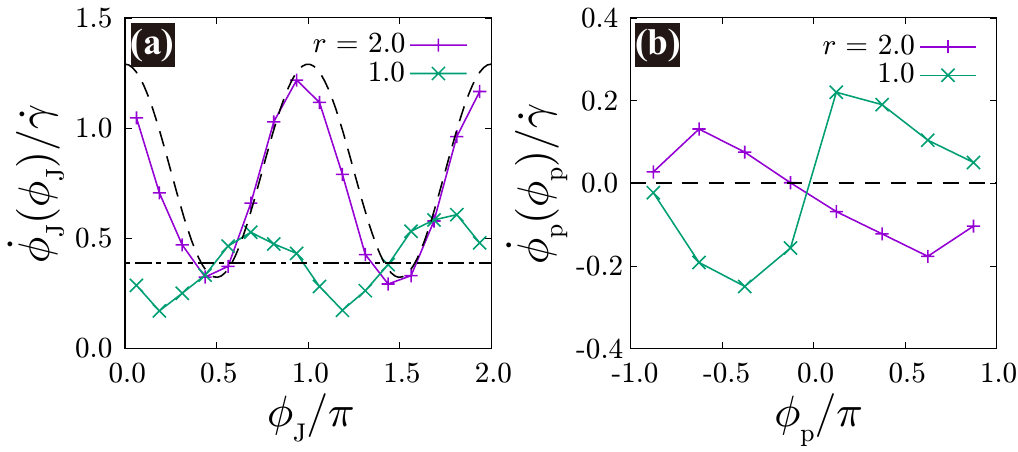}
  \caption{(Color online)
  (a) $\dot{\phi}_{\rm J}(\phi_{\rm J})$ in the bulk region ($|z/H|<0.2$) for $r=2.0$ and $1.0$ at $\phi=0.022$, $\dot\gamma=10^{-3}$ and $H=128$.
  For rod-like swimmers($r=2.0$), $\dot{\phi}_{\rm J}(\phi_{\rm J})$ closely follows Jeffery equation: $\dot{\phi}_{\rm J}(\phi_{\rm J}) = (\dot{\gamma}_{\infty}/(r^2+1))[r^2 \cos^2 \phi_{\rm J} + \sin^2 \phi_{\rm J}]$, where $\dot{\gamma}_{\infty} = 1.29 \times 10^{-3}$ is the average velocity gradient in the bulk region ($|z/H|<0.2$), while for spherical swimmers($r=1.0$), $\dot{\phi}_{\rm J}(\phi_{\rm J})$ fluctuates around $\dot{\gamma}_{\infty}/2$, $\dot{\gamma}_{\infty} = 0.78 \times 10^{-3}$ estimated in the bulk region. In the bulk, the shear rate is enhanced for rod-like swimmers and reduced for spherical swimmers, contributing to viscosity enhancement and reduction, respectively. For further details, please refer to Appendix B.
  (b) $\dot{\phi}_{\rm p}(\phi_{\rm p})$ near the top wall ($z/H>0.35$) for $r=2.0$ and $1.0$ under the same conditions as in (a).}
  \label{fig:phidot}
\end{figure}

Before proceeding, we here examine the rotational properties that are helpful for understanding the $H$-dependence of the swimming direction.
We first consider the rotational behavior in the $xz$-plane, characterized by   
\begin{eqnarray}
\dot{\phi}_{\rm J}(\phi_{\rm J}) & = & \frac{1}{N_{\rm b}}\sideset{}{^{\rm b}}\sum_{\alpha=1}^N \langle \dot{\phi}_{{\rm J},\alpha}(t) \delta[\phi_{\rm J} - \phi_{{\rm J},\alpha}(t)] \rangle,
\end{eqnarray}
where $\phi_{{\rm J},\alpha}$ ($0 < \phi_{{\rm J},\alpha} < 2 \pi$) is the azimuthal angle with respect to the $xz$-plain of the $\alpha$-th swimmer's orientation. 
The summation $\sum^{\rm b}$ indicates that we only consider swimmers in the bulk region ($|z/H|<0.2$), and $N_{\rm b}$ is the average number of the swimmers in this region.
Figure \ref{fig:phidot}(a) presents $\dot{\phi}_{\rm J}(\phi_{\rm J})$ for $r=2.0$ and $1.0$ at $\phi=0.022$, $\dot\gamma=10^{-3}$ and $H=128$.
For the rod-like swimmers ($r=2.0$), $\dot{\phi}_{\rm J}(\phi_{\rm J})$ closely follows the behaviors predicted by Jeffery's equation \cite{GrahamB,LaugaB}.
In contrast, for spherical swimmers ($r=1.0$), we observe $\dot{\phi}_{\rm J}(\phi_{\rm J}) \cong \dot{\gamma}_{\infty}/2$.
However, it exhibits a small periodic modulation, which induces a slightly nonzero value of $W_{xz}(z)$ in the middle region, as shown in Fig. \ref{fig:8moments}(g) \cite{discussion}.
As shown in Fig.  \ref{fig:phidot}(a), rotational dynamics in the $xz$-plane is significantly faster for rod-like swimmers than spherical ones, as expected, due to their larger sensitivity to torques.
Next, we analyze the rotational behavior in the $xy$-plane, described by 
\begin{eqnarray}
\dot{\phi}_{\rm p}(\phi_{\rm p}) & = & \frac{1}{N_{\rm tw}}\sideset{}{^{\rm tw}}\sum_{\alpha=1}^N \langle \dot{\phi}_{{\rm p},\alpha}(t) \delta[\phi_{\rm p} - \phi_{{\rm p},\alpha}(t)] \rangle,
\end{eqnarray}
where $\phi_{{\rm p},\alpha}$ ($-\pi < \phi_{{\rm p},\alpha} < \pi$) is the azimuthal angle with respect to the $xy$-plain. The summation $\sum^{\rm tw}$ considers only swimmers near the top wall for $Z_{\alpha}/H> 0.35$ are considered, and $N_{\rm tw}$ is the average number of the swimmers in this region.
Figure \ref{fig:phidot}(b) shows $\dot{\phi}_{\rm p}(\phi_{\rm p})$ for $r=2.0$ and $1.0$ at the same conditions as (a).
For rod-like swimmers, $\dot{\phi}_{\rm p}(\phi_{\rm p})$ is positive for $\phi_{\rm p} < 0$ and negative for $\phi_{\rm p} > 0$, indicating a tendency to align with the forward-flow direction.
This behavior can be attributed to the weather-vane effect \cite{Miki}.
In contrast, spherical swimmers exhibit the opposite tendency.
Unlike rod-like swimmers, the weather-vane effect is absent due to their isotropic body shape. Instead, as discussed in subsection IIIB, spherical swimmers exhibit a stronger vertical alignment tendency: once their swimming direction is established, this alignment becomes highly stable, preventing reorientation.
This point will be argued in some detail at the end of this subsection.

\begin{figure}[htb]
  \includegraphics[width=0.48\textwidth]{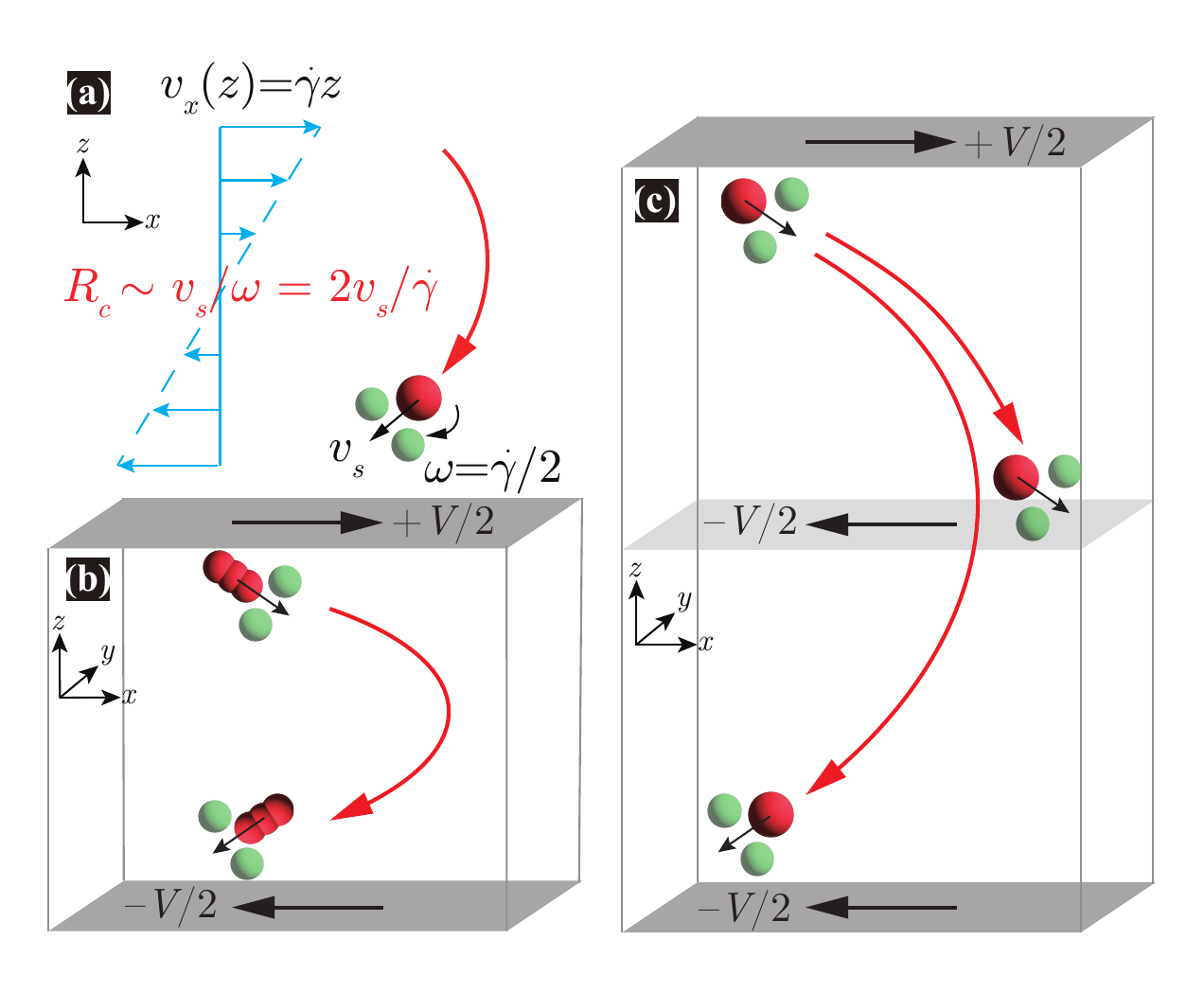}
  \caption{(Color online) 
  Schematic illustrations of a swimmer moving under shear-flow conditions. (a) In the bulk region, a swimmer typically moves while rotating clockwise, forming a cycloidal trajectory with a radius $R_{\rm c}$. 
  (b) For a rod-like swimmer under confinement, significant rotation can occur as it travels between the walls, even at smaller $H$. 
  (c) In contrast, a spherical swimmer requires a larger system height $H$ to make significant reorientation. The difference between these cases (b) and (c) arises from the difference in the hydrodynamic torque induced by the shear flow between rod-like and spherical swimmers.
  }
  \label{fig:cycloid}
\end{figure}

\begin{figure}[htb] 
  \includegraphics[width=0.48\textwidth]{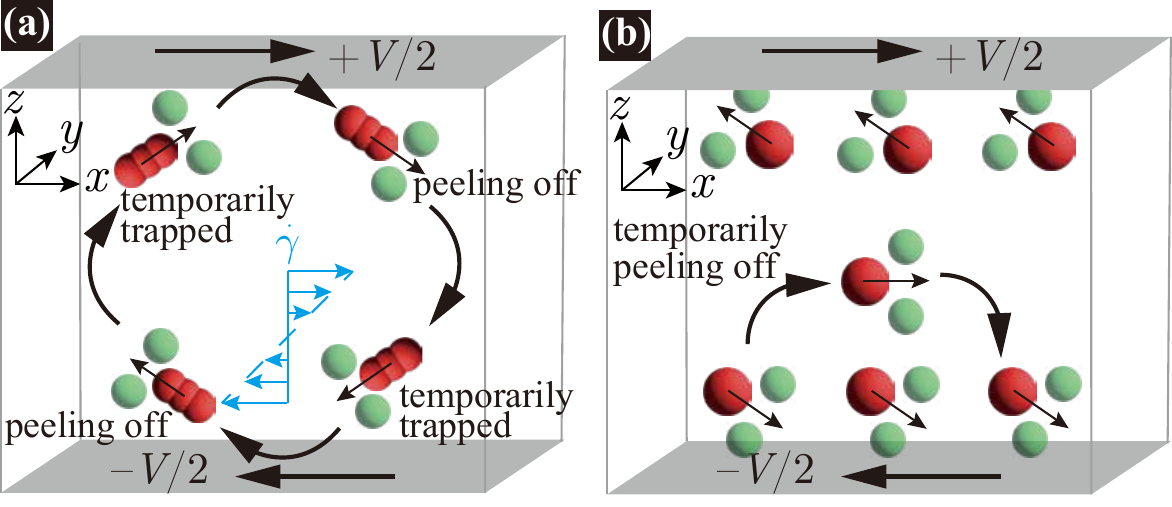}
  \caption{(Color online) 
  Schematic illustrations of distinct steady states:  
  (a) For the rod-like swimmers at $H$ above a certain threshold value, they can fully rotate clockwise, allowing them to establish a steady state where $\hat n_{\alpha,x} v_x>0$.  
  (b) For the spherical swimmers at $H$ below a certain threshold, once they begin moving against the flow direction ($\hat n_{\alpha,x} v_x<0$), this state tends to be maintained for a longer duration.
  }
  \label{fig:global_flow}
\end{figure}

Here, we examine the mechanisms underlying the strong $H$-dependence of the swimming direction. In our simulations, swimmers experience a shear-induced torque that rotates them, resulting in a clockwise trajectory, as schematically illustrated in Fig. \ref{fig:cycloid}(a). 
Let us consider a single spherical swimmer moving at a nearly constant speed $v_{\rm s}$ while rotating with an angular velocity of $\omega\sim \dot\gamma/2$ under a shear rate $\dot\gamma$. 
In an infinite bulk system, the trajectory forms a cycloid \cite{Hagen}: with $\omega\cong \dot\gamma/2=5\times 10^{-4}$ and $v_{\rm s}\cong 5\times 10^{-2}$, the characteristic cycloid radius is $R_{\rm c}=2v_{\rm s}/\dot\gamma\cong 10^2$.
When $H\lesssim 2R_{\rm c}$, spherical swimmers cannot rotate sufficiently between the walls, as illustrated schematically in Fig. \ref{fig:cycloid}(c). 
In contrast, as shown in Fig. \ref{fig:phidot}(a), rod-like swimmers are more strongly affected by the shear-induced torque than spherical swimmers, resulting in a higher angular velocity.
This enhanced rotational response enables rod-like swimmers to reorient more rapidly, even at relatively small $H$, as schematically illustrated in Fig. \ref{fig:cycloid}(b).
Consequently, for sufficiently large confinement height, a swimmer detaching from one wall while moving along the mean-flow direction ($\hat n_{\alpha,x} v_x>0$) can fully reorient during its travel and reach the opposite wall with satisfying $\hat n_{\alpha,x} v_x>0$.
This results in a steady global swimming state, where swimmers move in a clockwise trajectory in the $xz$-plane on average, as schematically illustrated in Fig. \ref{fig:global_flow}(a).

\begin{figure}[htb] 
  \includegraphics[width=0.25\textwidth]{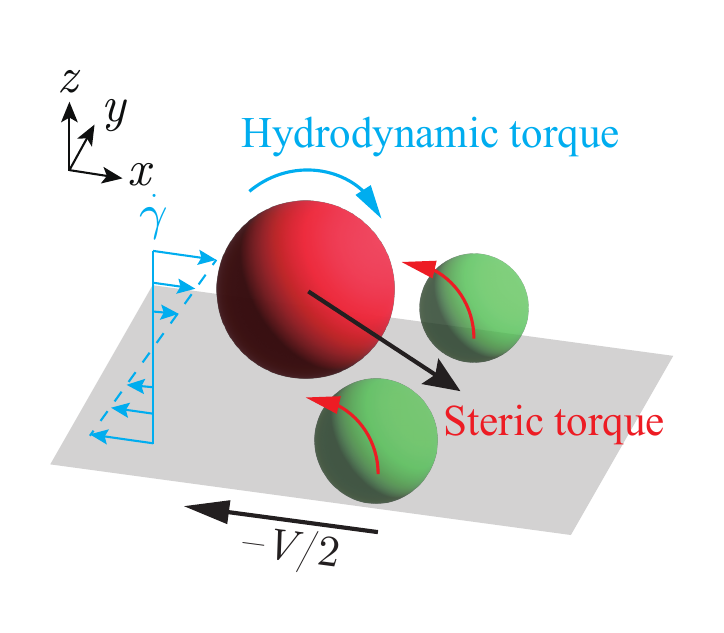}
  \caption{(Color online) A schematic illustration showing that when a swimmer moves along the backward-flow direction near the bottom wall, it experiences a clockwise rotational torque. However, steric interactions between its flagella and the wall hinder this rotational motion. }
  \label{fig:trapped_at_wall}
\end{figure}

Finally, we discuss the distinctive swimming behaviors observed in spherical swimmers at smaller confinement height $H$, considering the interplay of several effects discussed in this paper. 
A schematic representation of these behaviors is provided in  Fig. \ref{fig:global_flow}(b).
In the present flow geometry, the preferred swimming directions differ between the upper ($z>0$) and lower ($z<0$) regions. Thus, as swimmers travel between the two walls, they must reorient to align with the preferred direction ($\hat n_{\alpha.x} v_x>0$, i.e., the forward-flow direction).
First, we consider spherical swimmers moving along the preferred direction. 
As described earlier, under the present simulation setup, these swimmers rotate clockwise while moving with an almost constant swimming speed. However, when the confinement height is small ($H\lesssim 2R_c$), they do not have the necessary space to complete their reorientation during their travel to the other wall, preventing them from realigning with the preferred direction ($\hat n_{\alpha,x} v_x>0$).
Once swimmers move along the backward flow direction ($\hat n_{\alpha,x} v_x<0$) on the walls. Due to the following reasons, they tend to maintain this moving orientation for a longer duration:
i) Although these swimmers are subject to clockwise rotation, the steric interactions between their flagella part and the wall hinder this rotational motion.  A schematic illustration of this effect is shown in Fig. \ref{fig:trapped_at_wall}.
ii) Unlike rod-like swimmers, which exhibit a weather-vane effect causing the swimmers to align with the flow, spherical swimmers lack this mechanism.
iii) Swimmers that detach from one wall while keeping $\hat n_{\alpha,x} v_x<0$ experience a shear-induced torque that tends to rotate them back toward the departure wall, further preventing reorientation.
iv) As discussed in Subsection IIIB, spherical swimmers exhibit a stronger vertical alignment tendency, leading to the formation of larger vertically aligned clusters.
Therefore, in these consequences, once swimmers are in the state of $\hat n_{\alpha,x} v_x<0$, they tend to remain in this orientation for an extended period, significantly impacting their overall swimming behavior in confined environments.

In this subsection, we have demonstrated the significant effects of confinement on swimmer behavior. At a smaller confinement height, the limited available space restricts the reorientation, an effect that is particularly pronounced for spherical swimmers.
This is due to their lower responsibility to applied torques, which arises from their smaller aspect ratio compared to rod-like swimmers. As discussed above, other factors, such as vertical alignment tendencies, wall interactions, and the presence or absence of the weather-vane effect, also contribute to the observed swimming behaviors. A comprehensive analysis of these additional factors is beyond the scope of the present study, but will be explored in more detail in future work.

\section{Concluding remarks} 
In this study, we investigated the steady-state swimming properties and anomalous viscosity of model puller-type microswimmers using direct hydrodynamic simulations. Our swimmer models were designed to replicate {\it Chlamydomonas} by adjusting the shapes and relative positions of the body and flagellar components. 
For sufficiently large confinement heights, the active stress contribution $\Delta \eta_{\rm a}$ to the overall viscosity becomes positive, consistent with experimental observations \cite{Rafai, Mussler}. This viscosity increase is attributed to the pronounced orientational order of swimmers, particularly near the walls, where they tend to align with the mean-flow direction. This behavior is in marked contrast to our previous simulations of pusher-type swimmers \cite{Hayano}, where bulk orientational order leads to a significant viscosity reduction.

A key characteristic of the present puller-type swimmers is their intrinsic tendency for vertical alignment, caused by HIs mediated through the contractile flow fields they generate. This vertical alignment leads to stronger polar and orientational order than that observed in pusher-type swimmers under similar conditions (see Fig. \ref{fig:8moments} vs. Fig. 3 in Ref. \cite{Hayano}), and is reminiscent of the polar order reported in studies on squirmers \cite{Evans,Alarcon}.
Additionally, we have demonstrated strong confinement effects that significantly alter steady-state properties. For large confinement heights, swimmers predominantly follow the mean-flow direction. However, when the confinement height falls below a threshold $H_{\rm c}$, the swimmers are unable to fully reorient during their transit between opposing wall regions, resulting in suppressed polar and orientational order. Interestingly, in this highly confined regime, spherical swimmers can become trapped in an orientation opposite to that observed for $H\gg H_{\rm c}$. Moreover, both the threshold value $H_{\rm c}$ and the overall swimming behavior in confined systems are strongly influenced by swimmer shape, particularly aspect ratio.

Our findings emphasize the crucial role of microswimmer characteristics, such as propulsion mechanisms and morphology, in determining the rheological properties of active suspensions. 
Future studies will explore the emergence of a great variety of collective behaviors, aiming to further elucidate the dynamic coupling between hydrodynamic interactions and confinement effects in active matter systems.

\begin{acknowledgments}
This work was supported by JSPS KAKENHI (Grants No. 20H05619) and JSPS Core-to-Core Program ``Advanced core-to-core network for the physics of self-organizing active matter'' (Grants No. JPJSCCA20230002) and JST SPRING (Grants No. JPMJSP2108).
\end{acknowledgments}

\appendix
\section{Distinct swimming properties of spherical swimmers ($r=1.0$) at $H=128$}
Reflecting the geometric symmetry of the present system, the orientation distribution should exhibit symmetry with respect to the $z=0$ plane.
For rod-like swimmers ($r=2.0$), each simulation run shows that the steady-state orientation distribution indeed maintains this symmetry.
However, as mentioned in the main text, for spherical swimmers ($r=1.0$), while the orientation distribution averaged over 12 independent ensembles displays symmetry (shown in Fig. \ref{fig:8moments}), significant symmetry breaking is observed in individual runs, even after the system seems to reach a steady state. Here, we discuss the mechanisms behind this apparent symmetry-breaking for spherical swimmers.

\begin{figure}[thb] 
  \includegraphics[width=0.48\textwidth]{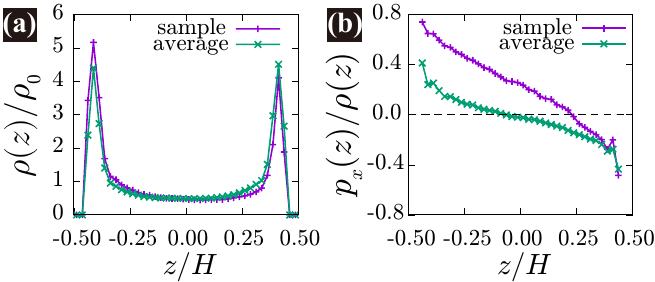}
  \caption{(Color online) 
  (a) The $z$-dependence of $\rho(z)$ for a sample run and the ensemble average for the spherical swimmers at $\phi=0.022$, $\dot\gamma=10^{-3}$ and $H=128$. (b) The $z$-dependence of $p_x(z)/\rho(z)$ at the same condition as (a).}
  \label{fig:asym}
\end{figure}

In Fig. \ref{fig:asym}, we plot the time-averaged profiles of $\rho(z)$ and $p_x(z)/\rho(z)$, both for a single ensemble and for the average over 12 independent ensembles. As shown in Fig. \ref{fig:asym}(a), $\rho(z)$ from a single run exhibits slight asymmetry with respect to the $z=0$ plane. This asymmetry is more pronounced in the ratio $p_x(z)/\rho(z)$, as demonstrated in Fig. \ref{fig:asym}(b). In the sample run shown here, there are more swimmers near the bottom boundary ($z/H=-0.5$) than near the top boundary ($z/H=0.5$), and the swimming directions of the swimmers at the more populated boundary dominate the overall distribution. It is worth noting that similar behavior is observed in other independent runs, although the boundary with a higher concentration of swimmers is randomly chosen.
Moreover, as the number of samples increases, the ensemble-averaged density and orientation distributions of the swimmers become more symmetric.

\begin{figure}[hbt]
  \includegraphics[width=0.48\textwidth]{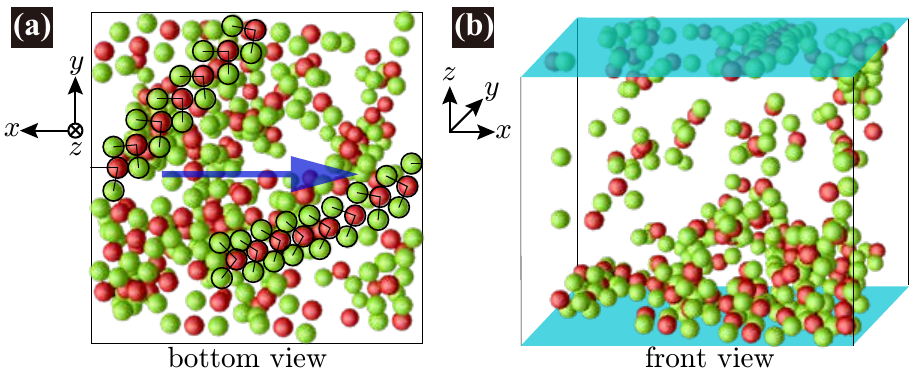}
  \caption{(Color online) 
  (a) The snapshot of the system for the spherical swimmers from the bottom view at $\phi=0.022$, $\dot\gamma=10^{-3}$ and $H=128$. The black circles and lines emphasize the swimmers that line up vertically. (b) The snapshot of the system from the front view at the same time step as (a).}
  \label{fig:snap_appendix}
\end{figure}

Typical snapshots from the current sample simulation run are shown in Fig. \ref{fig:snap_appendix}.
As displayed in Fig. \ref{fig:snap_appendix}(b), more swimmers are concentrated near the bottom boundary in this sample run.
As seen in the bottom view (Fig. \ref{fig:snap_appendix}(a)), the swimmers are aligned vertically along the walls. 
For puller swimmers, significant inflows toward the swimmer are generated both in front of and behind it, creating attractive HIs that cause the swimmers to align vertically along their orientation axes.
This effect is more pronounced near the walls, where the swimmers spend more time compared to the bulk region.
Due to the current setup of the spherical swimmers, once stacked structures form as a result of attractive HIs, they become sterically stable, grow in size, and persist for a time period comparable to the entire simulation duration. 
Note, however, that while rod-like swimmers also experience hydrodynamic attraction, their vertically aligned structures are less stable than those of spherical swimmers, due to greater transverse fluctuations.
Because of the large-scale collective motions of spherical swimmers, the wall with the higher concentration of swimmers remains nearly fixed throughout each simulation run, even though it is randomly determined.
That is, while the swimming properties averaged over many ensembles are consistent with the spatial symmetry imposed by the flow and system studied here, effective symmetry breaking can occur in a single independent run.

Finally, we note the following: preliminary simulations conducted with a doubled system size in the lateral dimensions ($L=256$), not shown here, indicate that system size effects in the lateral ($x$- and $y$-) directions remain qualitatively unchanged. This is in marked contrast to the strong confinement effects observed along the vertical direction. For spherical swimmers ($r=1.0$), they still tend to form elongated clusters along the walls. However, at $L=256$, these clusters span the entire system size less frequently compared to $L=128$.
Additionally, although HIs are inherently long-ranged, the system size used in the main analysis ($L=128$), combined with periodic boundary conditions, is sufficiently large to ensure that interactions with mirror-image swimmers do not dominate over those with real surrounding swimmers.

\section{Spatial modulation of velocity field}

\begin{figure}[tbh]
  \includegraphics[width=0.48\textwidth]{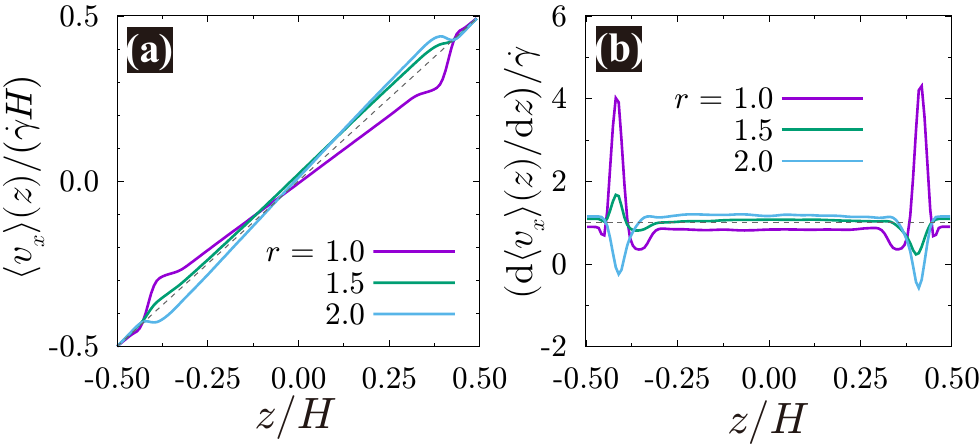}
  \caption{(Color online)
  (a) $\langle v_x\rangle(z)$, averaged over the $xy$-plane for $r=1.0, 1.5$ and $2.0$ at $\phi=0.022$, $\dot\gamma = 10^{-3}$, and $H=128$. (b) ${\rm d}\langle v_x\rangle(z)/{\rm d}z$ at the same condition as (a).
  }
  \label{fig:solvents}
\end{figure}

\begin{figure}[tbh] 
  \includegraphics[width=0.48\textwidth]{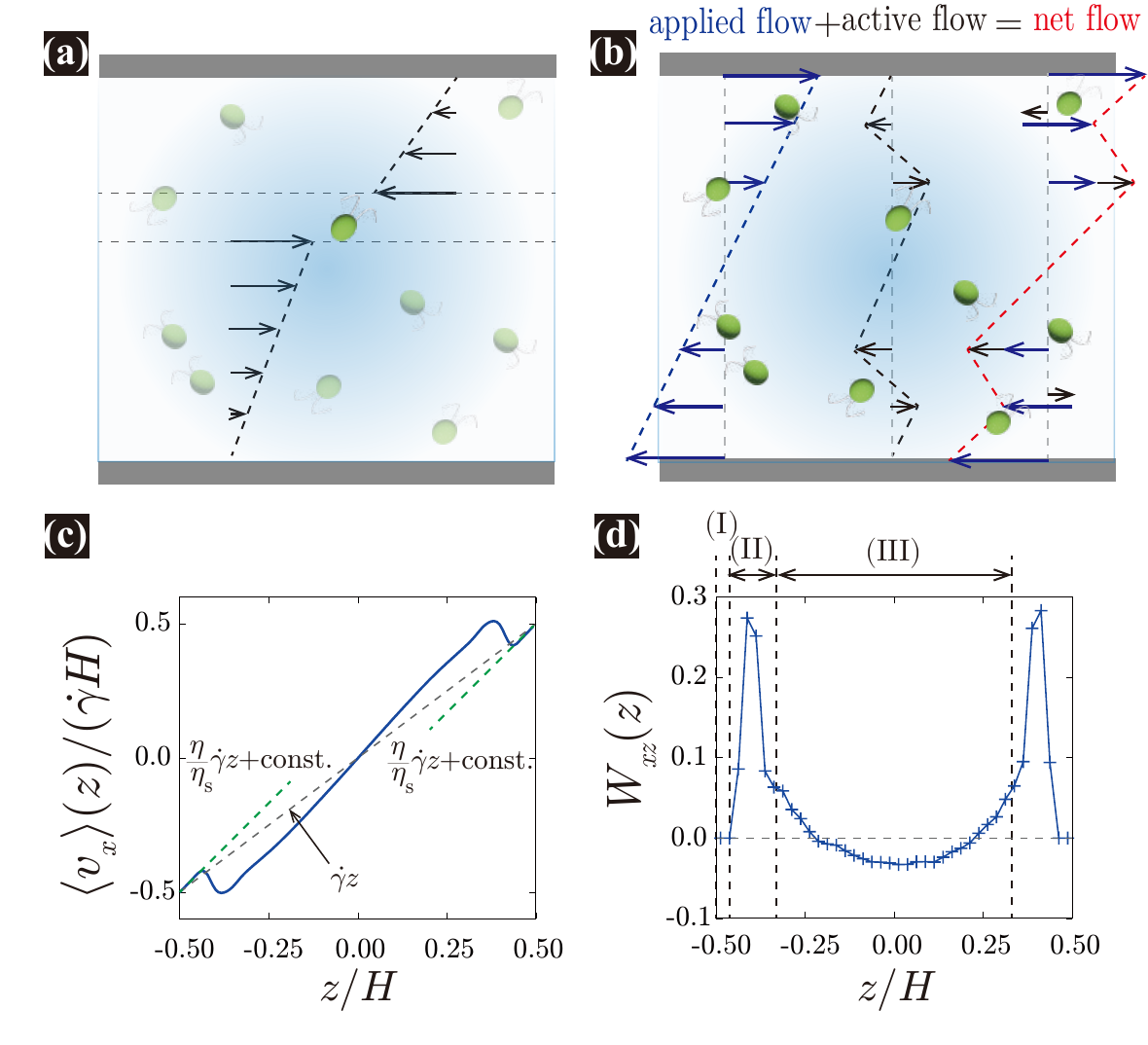}
  \caption{(Color online)
  (a) For a swimmer, when ${\hat n}_{\alpha,x} {\hat n}_{\alpha,z}>0$, the surrounding velocity gradient is weakened, while away from the swimmer, the opposite occurs.  (b) The net flow is determined by a superposition of the individual swimmers' contributions. $\langle v_x\rangle(z)$ (c) and $W_{xz}(z)$ (d), at $\phi=0.044$, $\dot\gamma = 10^{-3}$, and $H=128$. Near the walls, there are regions where a negative velocity gradient is observed, which closely corresponds to the region denoted as (II) [in (d)], where $W_{xz}(z)$ shows significant positive peaks. To compensate for these negative velocity gradients in region (II), the velocity field is accelerated in regions (I) and (III).
  }
  \label{fig:modulated_flow}
\end{figure}

In Fig. \ref{fig:solvents}(a), the $x$-component of the steady-state velocity field averaged over the $xy$-plane, $\langle v_x\rangle(z)$, is shown for $r=1.0$, $1.5$, and $2.0$ at $\phi=0.022$.  
Additionally, ${\rm d}\langle v_x\rangle(z)/{\rm d}z$, which is identical to the $z$-dependent solvent stress, is illustrated in Fig. \ref{fig:solvents}(b).
In the bulk region, the velocity gradient is enhanced and reduced at $r=2.0$ and $1.0$, respectively. 
On the other hand, near the walls, significantly negative velocity gradients and downward peaks in the solvent stress are observed for $r=2.0$, while the opposite occurs for $r=1.0$.
Note that, in both the bulk and wall regions, at $r=1.5$, the observed change in ${\rm d}\langle v_x\rangle(z)/{\rm d}z$ is much smaller than those at $r=2.0$ and $1.0$.
The viscosity observed at the walls results purely from solvent velocity gradients because the swimmers are not located in the narrow region near the walls due to the repulsive forces from the walls. 
Thus, this enhanced (reduced) viscosity immediately indicates the accelerated (decelerated) shear flow at the walls. 
In Figs. \ref{fig:modulated_flow}(a) and (b), we schematically display how the swimming state affects solvent flow for the case of $r=2.0$.
The following points are worth noting.
In the present puller systems, significant orientational order is more apparent around the walls than in the bulk regions. Consequently, steady-state swimming properties near the walls predominantly determine the resultant viscosity [see Eq. (\ref{eq:active_viscosity2})]. However, this tendency is completely different from the pusher case \cite{Hayano}, where the swimming properties in the bulk region control rheology for similar situations. This difference can be attributed to the difference in swimming mechanisms, specifically to the induced velocity fields, between pusher and puller swimmers.


\begin{thebibliography}{170}
\bibitem{comment_flow}
In Refs. \cite{Liu, Nash, Hill, Costanzo, Kaya}, a Poiseuille flow geometry is used, where swimmers move upstream. 
In the simple shear flow considered in this study, this corresponds to swimming along the forward-flow direction.
\bibitem{discussion}
A similar effect was reported in a previous study on pusher swimmers \cite{Hayano}, where hydrodynamic scattering under shear flow leads to an enhanced swimmer distribution along the extensional direction and a reduction along the compression direction. However, in the case of puller suspensions, the underlying mechanism of the observed anisotropy may differ from that of pusher suspensions and remains elusive.
\end{thebibliography}
\end{document}